\documentclass[amsthm,10pt]{elsart}
\usepackage{amsmath,amsthm,amsfonts,amssymb}
\usepackage[dvips]{graphicx}

\begin{document}

\begin{frontmatter}

\title{Domain decomposition improvement of quark propagator estimation}

\author{Tommy Burch}, 
\author{Christian Hagen}

\address{Institut f\"ur Theoretische Physik, Universit\"at 
Regensburg, D-93040 Regensburg, Germany}

\begin{abstract}
Applying domain decomposition to the lattice Dirac operator 
and the associated quark propagator, we arrive at expressions which, 
with the proper insertion of random sources therein, can provide 
improvement to the estimation of the propagator. 
Schemes are presented for both open and closed (or loop) propagators. 
In the end, our technique for improving open contributions is similar 
to the ``maximal variance reduction'' approach of Michael and Peisa, 
but contains the advantage, especially for improved actions, of 
dealing directly with the Dirac operator. 
Using these improved open propagators for the Chirally Improved 
operator, we present preliminary results for the static-light meson 
spectrum. 
The improvement of closed propagators is modest: 
on some configurations there are signs of significant noise reduction 
of disconnected correlators; 
on others, the improvement amounts to a smoothening of the same 
correlators.
\end{abstract}


\begin{keyword}
Lattice gauge theory, hadron spectroscopy
\end{keyword}

\end{frontmatter}

\section{Introduction}\label{SectIntroduction}

In recent years, there has been a number of techniques developed for 
improving estimates of quark propagators on the lattice (see, e.g., 
Refs.\ \cite{Michael:1998sg,Duncan:2001ta,Bali:2005fu,Foley:2005ac}). 
The main idea is to try to devise an estimator formulation which allows one 
to use all (or at least many) lattice sites as source locations for the 
quarks, rather than just having a fixed location, which is traditionally the 
case. 
If successful, one then has an estimate of the quark propagators from 
anywhere to anywhere in the lattice (typically called ``all-to-all'' 
propagators).

In the following sections we add our voices to the crowd and present our 
own method, which is based, foremostly, upon domain decomposition, or more 
specifically, the decomposition of block matrices. 
We end up with two different types of estimators: one for ``open'' 
propagators between two domains of the lattice (or ``half-to-half'') and 
one for ``closed'' propagators within one of the domains.

We first show how we devised our method through the consideration of random 
sources. 
Later we present first results for the two types of estimators and some 
of their possible applications: 
static-light mesons for the open propagators and disconnected correlators 
for the closed.

\section{The method}\label{SectMethod}

Using a set of random sources, $\chi_{j \beta b}^n$, placed at all sites of 
the lattice, one can determine a corresponding set of solution vectors 
(we use a summation convention for repeated indices throughout this paper): 
\begin{equation}
  \eta_{i \alpha a}^n = M_{i \alpha a \, j \beta b}^{-1} \: \chi_{j \beta b}^n \;\;\; (n=1,\ldots,N) \, .
\end{equation}
Since we know that 
$\lim_{N\to\infty} \frac1N \chi_{j \beta b}^n \chi_{k \gamma c}^{n\dagger} = \delta_{jk} \delta_{\beta \gamma} \delta_{bc}$, 
we can construct an estimate of the full propagator: 
\begin{equation}
  \label{naive_estimator}
  M_{i \alpha a \, k \gamma c}^{-1} \approx \frac1N \eta_{i \alpha a}^n \chi_{k \gamma c}^{n\dagger} \, .
\end{equation}
This is, however, a rather noisy estimator and we map out some improvement 
schemes in the following sections.

The main feature of our approach involves the consideration of independent 
regions of the lattice, or domain decomposition, a technique used previously 
for ``maximal variance reduction'' (MVR) of pseudofermion estimators 
\cite{Michael:1998sg}. Similar considerations may also be used for improving 
dynamical lattice updates \cite{Luscher:2005rx}.

We can think of the lattice as a disjoint union of two regions. 
The full Dirac matrix can then be written in terms of submatrices 
\begin{equation}
  M = \left(
    \begin{array}{cc}
      M_{11} & M_{12} \\
      M_{21} & M_{22}
    \end{array} \right) \, ,
\end{equation}
where $M_{11}$ and $M_{22}$ connect sites within a region and 
$M_{12}$ and $M_{21}$ connect sites from the different regions. 
(For simplicity, we suppress color, spin, and site indices in the following.) 
Regardless of the shape or 
nature\footnote{For example, different regions in color or Dirac space.} 
of the regions, a similarity transformation is all that is needed to reach 
this form. 
We can also write the propagator in this form: 
\begin{equation}
  M^{-1} = P = \left(
    \begin{array}{cc}
      P_{11} & P_{12} \\
      P_{21} & P_{22}
    \end{array} \right) \, .
\end{equation}

\subsection{Open contributions}

It is helpful to imagine reconstructing the sources in one region, $\chi_1^n$, 
from the solution vectors everywhere, $\eta^n$, and to separate the 
contributions from the different regions: 
\begin{equation}
  \chi_1^n = M \eta^n = M_{11} \eta_1^n + M_{12} \eta_2^n \, .
\end{equation}
If we now apply the inverse of the matrix within one region, we have 
\begin{equation}
  M_{11}^{-1} \chi_1^n = \eta_1^n + M_{11}^{-1} M_{12} \eta_2^n \, .
\end{equation}
This can be solved for $\eta_1^n$ and substituted into the 
original expression for the noisy estimator of the propagator between the 
two regions: 
\begin{eqnarray}
  \left( M^{-1} \right)_{12} = 
  P_{12} &\approx& \frac1N \eta_1^n \chi_2^{n\dagger} \nonumber \\
  &\approx& \frac1N \left[ M_{11}^{-1} \left( \chi_1^n - M_{12} \eta_2^n \right) \right] \chi_2^{n\dagger} \nonumber \\
  &\approx& - \frac1N \left( M_{11}^{-1} M_{12} \eta_2^n \right) \chi_2^{n\dagger} \, ,
\end{eqnarray}
where in the last line we eliminate the first term due to the fact that 
we expect $\lim_{N\to\infty}\chi_1^n \chi_2^{n\dagger} = 0$. 
This is a crucial step, for here we cut out of the calculation what 
amounts to being only noise. It does not come for free, however, since we 
must perform the additional inversion within the subvolume $1$. 
Writing out the full expression, we obtain 
\begin{eqnarray}
  \label{postdiluting}
  P_{12} &\approx& - \frac1N M_{11}^{-1} M_{12} P \chi_2^n \chi_2^{n\dagger} \nonumber \\
  &=& - M_{11}^{-1} M_{12} P_{22} \, ,
\end{eqnarray}
where the second line is an exact expression, showing that one 
can relate elements of different regions of $P=M^{-1}$ via the inverse of a 
submatrix of $M$. 
(We do not pretend to have derived something new here; after all, 
$P_{22}$ is the Schur complement of $M_{11}^{-1}$. 
We only wish to emphasize the useful connection with random-source 
techniques.) 
Again, the lesson learned up to this point is that we need no sources in 
one of the two regions. 
The story does not end here, however.

Looking again at Eq.\ (\ref{postdiluting}), one can see that we need not make 
the approximation 
$P_{22} \approx \frac1N P \chi_2^n \chi_2^{n\dagger}$. 
Instead, we can place the approximate Kronecker delta between the $M_{12}$ 
and $P_{22}$: 
\begin{eqnarray}
  \label{conn_estimator}
  P_{12} &\approx& - \frac1N M_{11}^{-1} M_{12} \chi_2^n \chi_2^{n\dagger} P \nonumber \\
  &\approx& - \frac1N \left( M_{11}^{-1} M_{12} \chi_2^n \right) \left( \gamma_5 P \gamma_5 \chi_2^{n} \right)^\dagger \nonumber \\
  &\approx& - \frac1N \psi_1^n \phi_2^{n\dagger} \, ,
\end{eqnarray}
where we have used the $\gamma_5$-hermiticity of the propagator 
(this is done only for convenience since we could just as well work with 
$P^\dagger\chi_2^n$ in the $\phi_2^n$). 
From the next to last line, one can see that the signal only rises from 
terms where the two components of the $\chi_2^n$'s are the same. 
However, unlike the naive estimator, Eq.\ (\ref{naive_estimator}), 
where there is only 1 such term giving a signal-to-noise of $\sim 1/V^{1/2}$, 
here there are many: for $V_\chi$ source points, the signal-to-noise is 
$\sim V_\chi / (V_\chi^2-V_\chi)^{1/2} \sim 1$. 
Terms where the components of the sources are not the same can still be 
eliminated by ``dilution'' of the original source vectors, $\chi^n$, that go 
into the $\psi_1^n$, the $\phi_2^n$, or for greater noise reduction, both 
simultaneously. 
Probably more important than these considerations, however, is the fact that, 
for most of the propagators between the regions, the random sources are kept 
far from the end points. 
Also, one can use all points in one region for the source and all points in 
the other region for the sink.

\begin{figure}
\begin{center}
\includegraphics*[width=10cm]{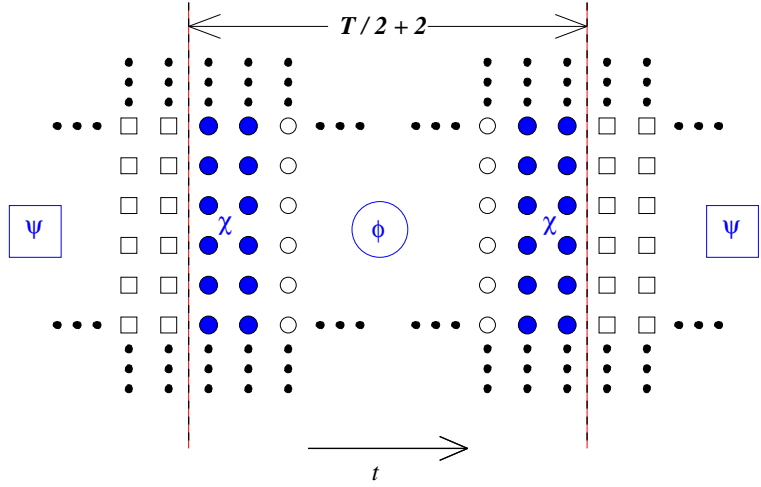}
\end{center}
\caption{
Depiction of the ideal decomposition and the sources ($\chi$) 
which are needed to construct the estimate of the CI quark propagator 
between regions.
}
\label{ideal_sources}
\end{figure}

The ideal domain decomposition for quark propagators which contribute to 
connected diagrams is then to use two unequal volumes, one containing a few 
more time slices (those of the sources $\chi$) than the other. Ideally, 
the centers of the two sets of source time slices should be separated by 
$T/2$. The number of source time slices is dictated by the lattice Dirac 
operator since the $\chi$'s should be placed on all time slices which 
communicate with the other region via one application of $M$. 
For Wilson and Fixed-Point (FP) \cite{FP_action} quarks, this is just 2 time 
slices, 1 adjacent to each boundary. For Chirally Improved (CI) quarks 
\cite{CI_action}, which we use here, 4 are necessary 
(see Fig.\ \ref{ideal_sources}). 
For the Asqtad action \cite{Asqtad_action}, 6 are needed due to the presence 
of the Naik term. 
While for Overlap fermions \cite{Overlap}, it might be best to 
use equal volumes for the two regions since the sources will have to cover 
one region entirely. 
But we stress that for all of the above, one is free to dilute further: e.g., 
by inverting the sources on the different time slices separately.

\begin{figure}
\begin{center}
\includegraphics*[width=8cm]{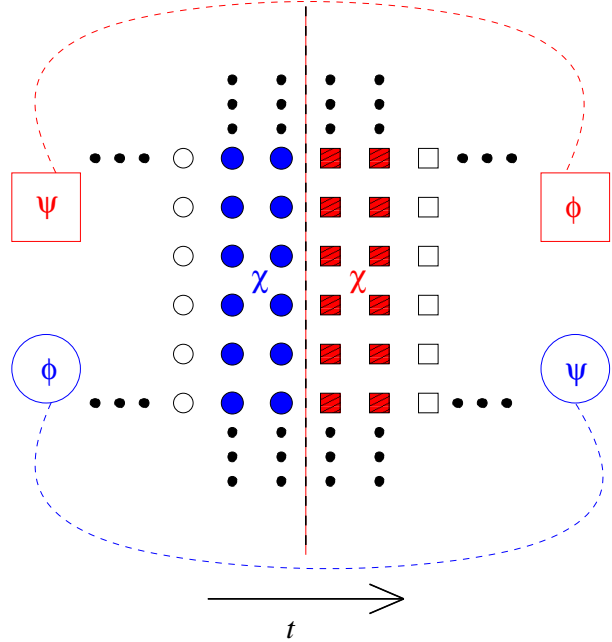}
\end{center}
\caption{
Depiction of one of the boundaries and the surrounding sources ($\chi$) 
which we use to construct the two estimates of the CI quark propagator 
between regions of equal volume. 
Colors and shapes indicate which source region contributes 
to the signal in the resultant vectors. 
The $\psi$'s are calculated using only one of the source regions, while 
the $\phi$'s use both.
}
\label{sources}
\end{figure}

For our first attempt of using this method, we do not choose 
the ideal decomposition. We use equal volumes for the two regions and 
place sources next to the boundaries in both regions 
(see Fig.\ \ref{sources}). Although this choice may not be ideal, we do 
perform inversions for each spin component separately (spin dilution). 
Note also that, since our sources occupy all relevant time slices surrounding 
the boundaries, we can actually obtain two independent estimates of the quark 
propagator between the two regions: 
\begin{equation}
  \label{conn_estimator2}
  - \frac1N \psi_1^n \phi_2^{n\dagger} \approx P_{12} \approx 
  - \frac1N \gamma_5 \phi_1^n \psi_2^{n\dagger} \gamma_5 \, .
\end{equation}

So our method is very similar to that of MVR, except 
for the fact that we can work directly with $M$, instead of $M^\dagger M$. 
This gives the current method three advantages: First, it is less 
problematic to invert $M$ since it has a better condition number than 
$M^\dagger M$ \cite{Michael:1998sg}. 
Second, implementing the method is straightforward. 
One needs only to limit the range of application of $M$ when performing 
inversions in the subvolumes and when multiplying by the matrix between 
regions. Otherwise, existing routines remain unchanged. 
Third, the sources need only occupy enough time slices to connect them 
with the other region via $M$, rather than $M^\dagger M$. 
These are the same number of sources for Wilson-like operators, where 
$M^\dagger M$, like $M$, only extends one time slice. However, for many 
other improved operators (like CI and FP) this can reduce the number of 
necessary source time slices by a factor of 2.

\subsection{Closed contributions}

Working with the propagator within one of the regions, say 1, and following 
a similar derivation as in the previous section, one obtains an 
expression potentially useful for estimating quark propagators which return 
to the same region: 
\begin{eqnarray}
  \label{postdiluting2}
  P_{11} &\approx& \frac1N M^{-1}_{11} \left( \chi^n \chi_1^{n\dagger} - M_{12} P \chi^n \chi_1^{n\dagger} \right) \nonumber \\
  &=& M^{-1}_{11} - M^{-1}_{11} M_{12} P_{21} \, .
\end{eqnarray}
So once again, through the consideration of random sources, we arrive at an 
exact expression (and again, one which is nothing new). 
Now, combining the expressions for $P_{11}$ and $P_{21}$ 
($=\gamma_5 P_{12}^\dagger \gamma_5$), we arrive at the relation: 
\begin{equation}
  \label{whoa_nellie}
  P_{11} = M^{-1}_{11} + M^{-1}_{11} M_{12} \gamma_5 \left( M_{11}^{-1} M_{12} P_{22} \right)^\dagger \gamma_5 \, .
\end{equation}
Inserting our random sources into this expression gives 
\begin{eqnarray}
  \label{disc_estimator2}
  P_{11} &\approx& M^{-1}_{11} + \frac1N \left( M^{-1}_{11} M_{12} \gamma_5 \chi_2^n \right) 
  \left( \gamma_5 M_{11}^{-1} M_{12} P \chi^n \right)^\dagger \nonumber \\
  &\approx& M^{-1}_{11} + \frac1N \zeta_1^n \theta_1^{n\dagger} \, .
\end{eqnarray}
Note that we put no region index on the second $\chi$, indicating that for 
this resultant vector ($\theta_1$) we wish to use sources initially placed 
everywhere on the lattice. 
The advantage of expression (\ref{disc_estimator2}) may not be immediately 
clear since it still contains the explicit appearance of $M^{-1}_{11}$.

\begin{figure}
\begin{center}
\includegraphics*[width=10cm]{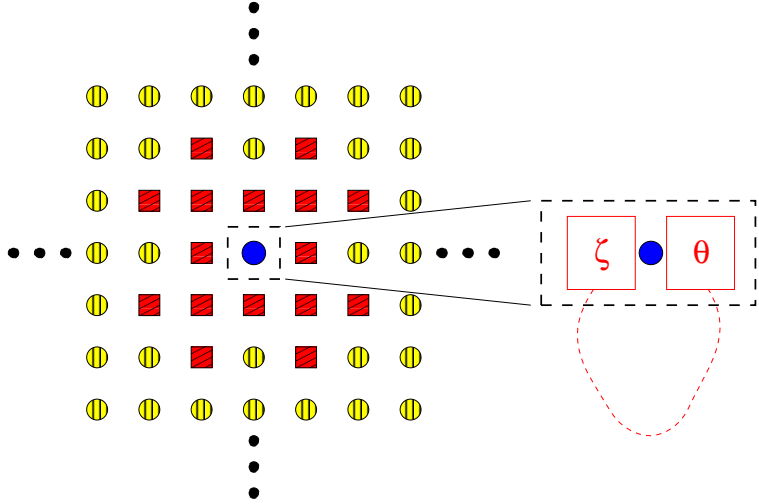}
\end{center}
\caption{
Depiction of the boundary and the surrounding region of sources which are 
needed to construct the estimated CI quark propagator to and from the 
central point. 
In this figure, region 1 is just one site (we also consider a larger region 
which includes nearest neighbors). 
The red squares indicate which region of sources contributes to the signal in 
the resultant vectors ($\zeta_1$, $\theta_1$) of Eq.\ (\ref{disc_estimator2}).
}
\label{disc_sources}
\end{figure}

This seeming hindrance can be remedied by considering a very small volume for 
region 1. Performing this ``highly reduced'' inversion exactly, we hope 
to find a significant gain in the signal-to-noise ratio: 
The first term in the above expressions will be exact and the second term, as 
compared to its naive estimate, may be improved by a factor of as much as 
$\sim V_\chi^{1/2}$, where $V_\chi$ is now the volume of sources in region 2 
which connect to region 1 via $M$. 
So if the volume of region 1 is kept small enough and the lattice Dirac 
operator connects each site to many others, there may be an advantage to 
calculating the $M^{-1}_{11}$'s exactly, as opposed to simply inverting more 
sources.

Here, we propose to use spin-diluted, but full-volume sources for $\chi$, so 
that we can, in the end, use all sites as the starting and ending point of 
the propagator $P_{xx}$ with only $N$ original full-volume inversions: 
the $P \chi^n$ in $\theta_1^n$. 
We can consider the smallest, symmetrical volume for region 1, 
the point itself ($V_1=1$; see Fig.\ \ref{disc_sources}), in order to reduce 
the amount of work needed to calculate $M^{-1}_{11}$, which we need for each 
point in the lattice. With this choice, at most one needs to invert 
$V$ $12\times12$ matrices. 
Since $V_\chi=128$ for the CI operator in this scenario, we hope 
that this small amount of extra work may be worth the effort. 
In the end, however, we actually use a larger region 1, including nearest 
neighbors ($V_1=9$), but we avoid some extraneous work by only inverting 
$M_{11}$ upon 12 sources located at the central point.

\section{First results}

In this section, we present some first results obtained with our estimated 
quark propagators. 
We first use the open propagators for the light quark in heavy-light meson 
correlators, where we use the static approximation for the heavy quark. 
We also show the effects of using our closed propagators in disconnected 
correlators.

\subsection{Static-light mesons}

For our meson source and sink operators, we use bilinears of 
the form: 
\begin{equation}
  \bar{Q} \, O(\Gamma,D_i,\vec D^2,S) \, q \, ,
\end{equation}
where $S$ is a gauge-covariant (Jacobi) smearing operator \cite{jacobi} and 
$\vec D$ is the covariant derivative. 
For our basis of light-quark spatial wavefunctions, we use three different 
amounts of smearing and apply 0, 1, and 2 covariant Laplacians to these: 
\begin{equation}
  q' = \; S_8 \, q \; , \; \vec D^2 S_{12} \, q \; , \; 
  \vec D^2 \vec D^2 S_{16} \, q \; ,
\end{equation}
where the subscript on the smearing operator denotes the number of smearing 
steps; all are applied with the same weighting factor of $\kappa_{sm}=0.2$. 
So we have a relatively narrow, approximately Gaussian distribution, along 
with wider versions which exhibit one and two radial nodes, due to the 
application of the Laplacians. 
We point out that, thus far, we have not altered the quantum numbers of the 
meson source since both the smearing and Laplacians treat all spatial 
directions the same (i.e., they are scalar operations).
In order to create mesons of different quantum numbers, we use these 
light-quark distributions together with the operators shown in 
Table \ref{sl_ops} (see, e.g., Ref.\ \cite{Michael:1998sg}).

\begin{table}
\caption{
Static-light meson operators.
\label{sl_ops}
}
\begin{tabular}{lcc} \hline
oper. & $J^P$ & $\bar{Q} \, O(\Gamma,D_i) \, q'$ \\ \hline
$S$ & $0^-,1^-$ & $\bar{Q} \, \gamma_5 \, q'$ \\
$P_-$ & $0^+,1^+$ & $\bar{Q} \, \sum_i \gamma_i D_i \, q'$ \\
$P_+$ & $1^+,2^+$ & $\bar{Q} \, (\gamma_1 D_1 - \gamma_2 D_2) \, q'$ \\
$D_\pm$ & $1^-,2^-,3^-$ & $\bar{Q} \, \gamma_5(D_1^2 - D_2^2) \, q'$ \\ \hline
\end{tabular}
\end{table}

Inserting the estimated and static propagators in the meson correlators we 
have 
\begin{eqnarray}
  C_{ij}(t) &=& \left\langle 0 \left| (\bar Q \, O_j \, q)_t \; 
      (\bar q \, \bar O_i \, Q)_0 \right| 0 \right\rangle \nonumber \\
  &=& \left\langle \sum_x \mbox{Tr} \left[ 
      \frac{1+\gamma_4}{2}\prod_{i=0}^{t-1} U_4^\dagger(x+i\hat{4}) \, 
      O_j P_{x+t\hat{4},x} \bar O_i 
    \right] \right\rangle_{\{U\}} \, .
\end{eqnarray}
The static quark is propagated through products of links in the time 
direction and has a fixed spin $(1+\gamma_4)/2$. The estimated propagator 
$P_{x+t\hat{4},x}$ is of the form of Eq.\ (\ref{conn_estimator2}). 
Thus, all points within region 1 ($N_s^3N_t/2$ of them) can act as the 
source location $x$, just so long as $t$ is large enough to have the sink 
location $x+t\hat{4}$ in region 2. 
Note that we now have subscripts on the source and sink operators to denote 
which light-quark distribution is being used. 
We create all such combinations and thus have a $3\times3$ matrix of 
correlators for each of the operators in Table \ref{sl_ops}.

Following the work of Michael \cite{Mi85} and 
L\"uscher and Wolff \cite{LuWo90}, 
we use this cross-correlator matrix in a variational approach to separate 
the different mass eigenstates. 
We must therefore solve the generalized eigenvalue problem 
\begin{equation}
  \label{gep}
  C(t) \vec v^{(k)} = \lambda^{(k)}(t,t_0) \, C(t_0) \vec v^{(k)} \; ,
\end{equation}
in order to obtain the following eigenvalues: 
\begin{equation}
  \label{eigenvalues}
  \lambda^{(k)}(t,t_0) \propto e^{-t \, M_k} 
  [1+O(e^{-t \, \Delta M_k})] \, ,
\end{equation}
where $M_k$ is the mass of the $k$th state and $\Delta M_k$ is the 
mass-difference to the next state. 
For large enough values of $t$, each eigenvalue should then correspond to a 
unique mass state, requiring only a single-exponential fit.

Before solving the above eigenvalue problem, we check that our 
cross-correlator matrix is real and symmetric (within errors). 
We then make it symmetric before solving Eq.\ (\ref{gep}) and look for 
regions of $t$ where the eigenvalues exhibit single masses and the 
corresponding eigenvectors, $\vec v^{(k)}$, are stable (each of these 
provides what we call the ``fingerprint'' of a state and if it holds steady, 
we have better reason to believe that we are looking at a single state). 
The results should be independent of the normalization time slice $t_0$ and 
we make sure that this is so. For our final results, however, we use the 
value of $t_0/a=1$.

This variational approach has seen much use recently in lattice QCD, 
especially for extracting excited hadron masses and we point the reader 
to the relevant literature in \cite{exc_had}.

We create our cross-correlator matrices on two sets of gluonic 
configurations: 100 quenched and 74 dynamical, each with $12^3\times24$ 
lattices sites. 
The quenched configurations have a lattice spacing of $a \approx 0.15$ fm 
($a^{-1}\approx1330$ MeV) and a spatial extent of $L\approx1.8$ fm. 
The dynamical set \cite{Lang:2005jz} has 2 flavors of CI sea quarks (with 
$M_{\pi,\mbox{sea}}\approx500$ MeV), $a \approx 0.115$ fm 
($a^{-1}\approx1710$ MeV), and $L\approx1.4$ fm. 
We use 12 random spin-color vectors as sources for the light-quark 
propagator estimation. Spin-diluted, this gives us 48 separate sources for 
the inversions (one in the full volume, $\phi$, and two in the subregions, 
$\psi$; see Eqs.\ (\ref{conn_estimator}) and (\ref{conn_estimator2})). 
We perform inversions for 4 different quark masses: 
$am_q=0.02$, 0.04, 0.08, 0.10.

\begin{figure}
\begin{center}
\includegraphics*[width=6.5cm]{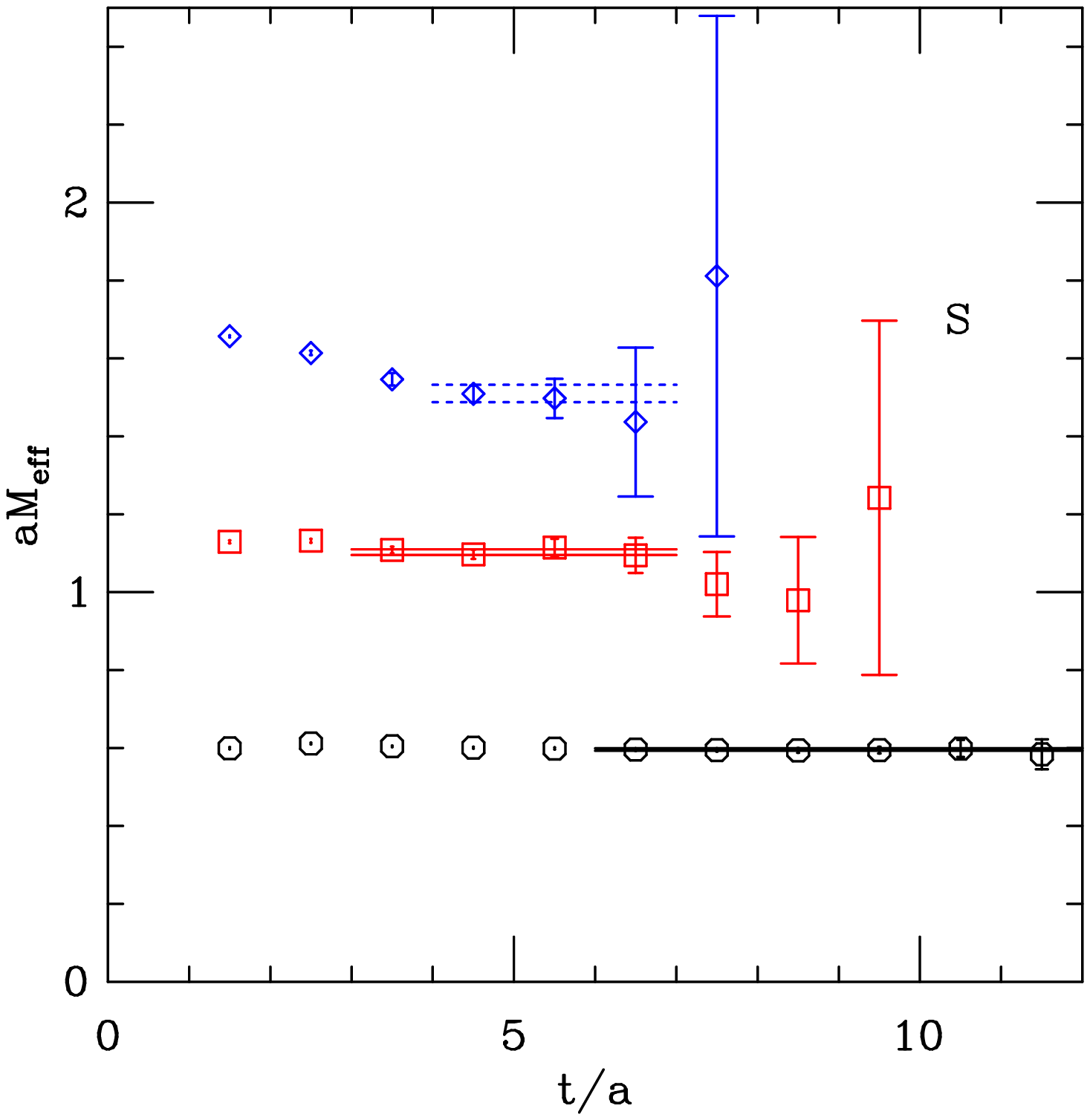}
\includegraphics*[width=6.5cm]{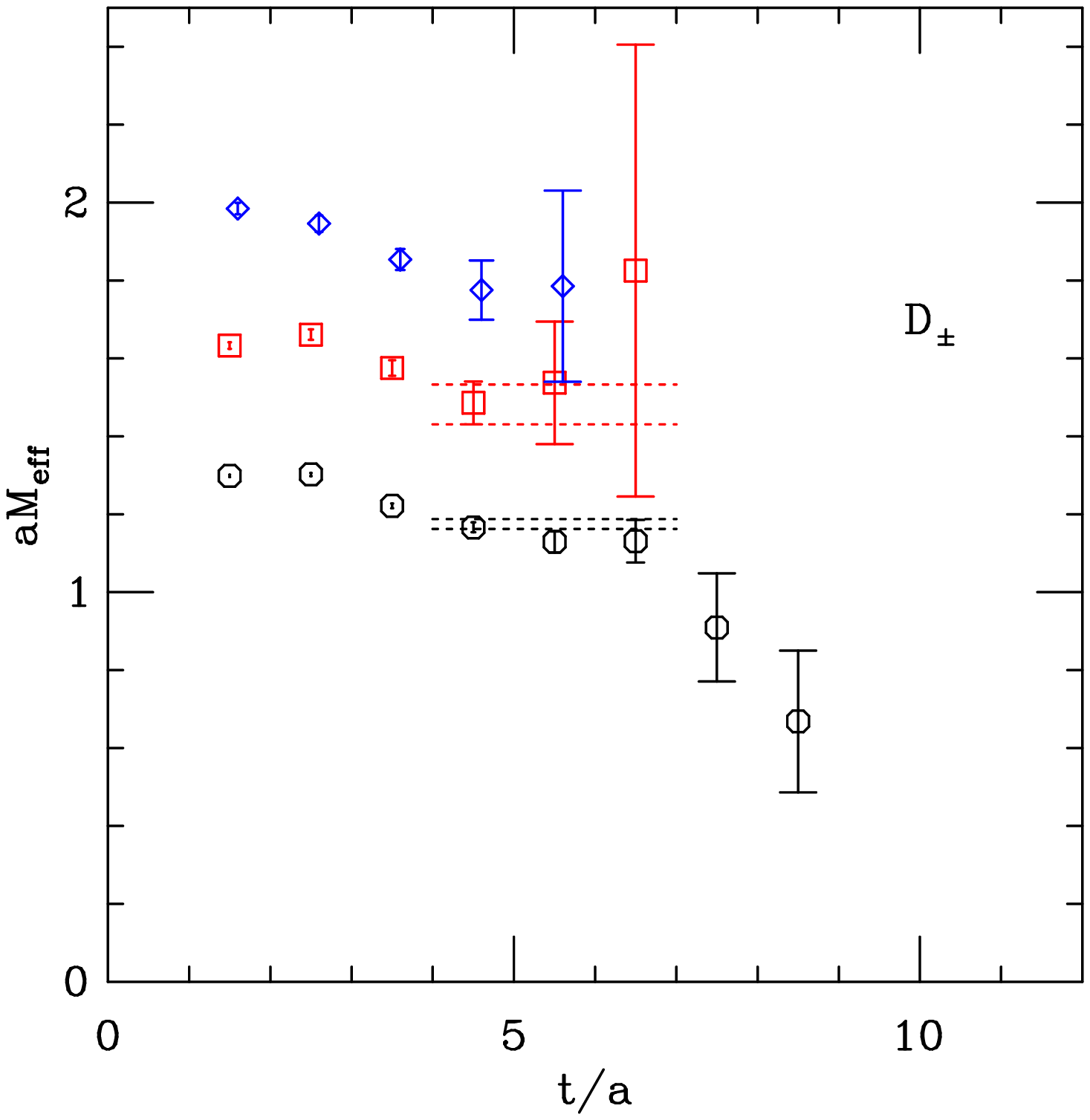}\\
\includegraphics*[width=6.5cm]{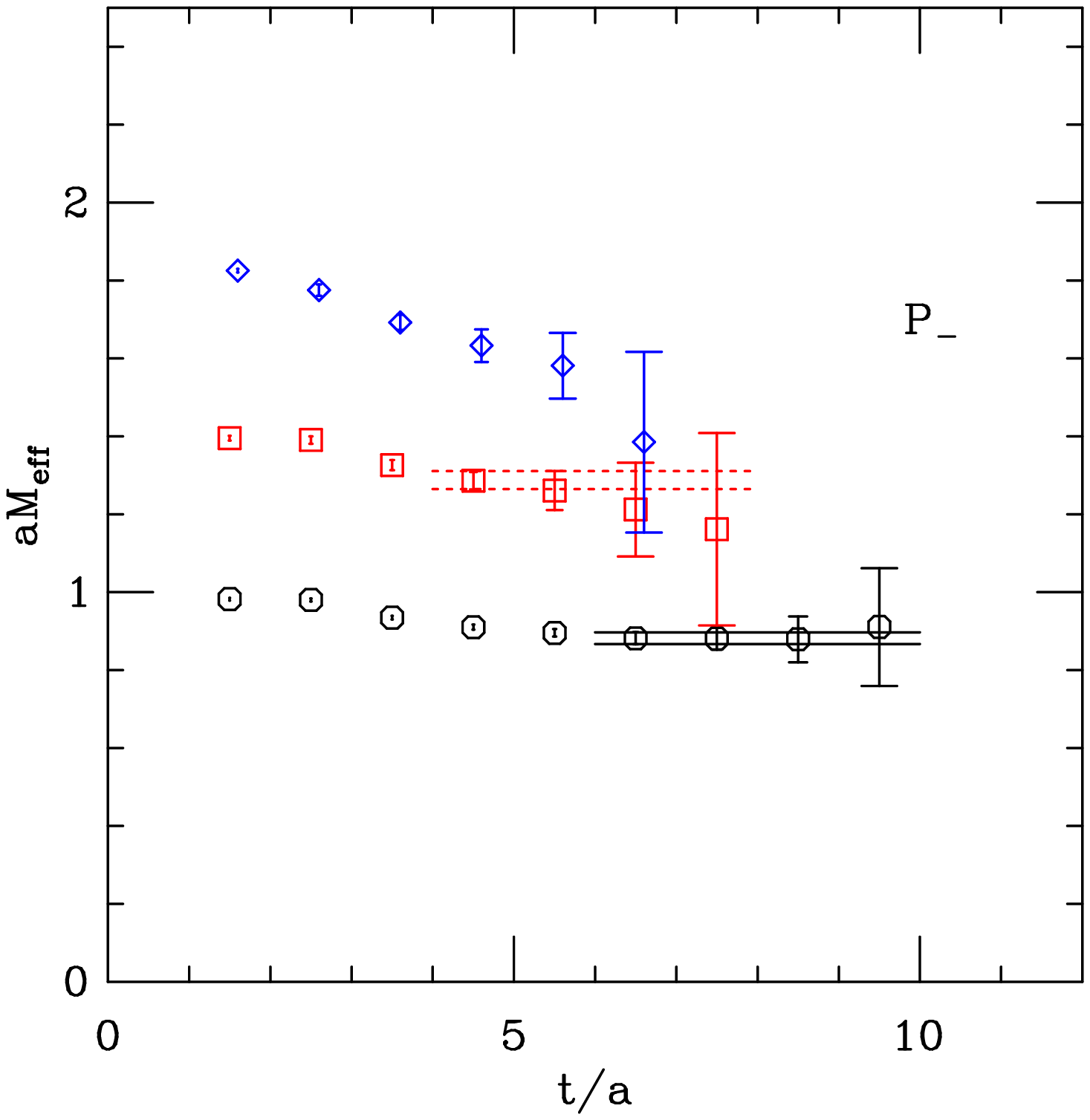}
\includegraphics*[width=6.5cm]{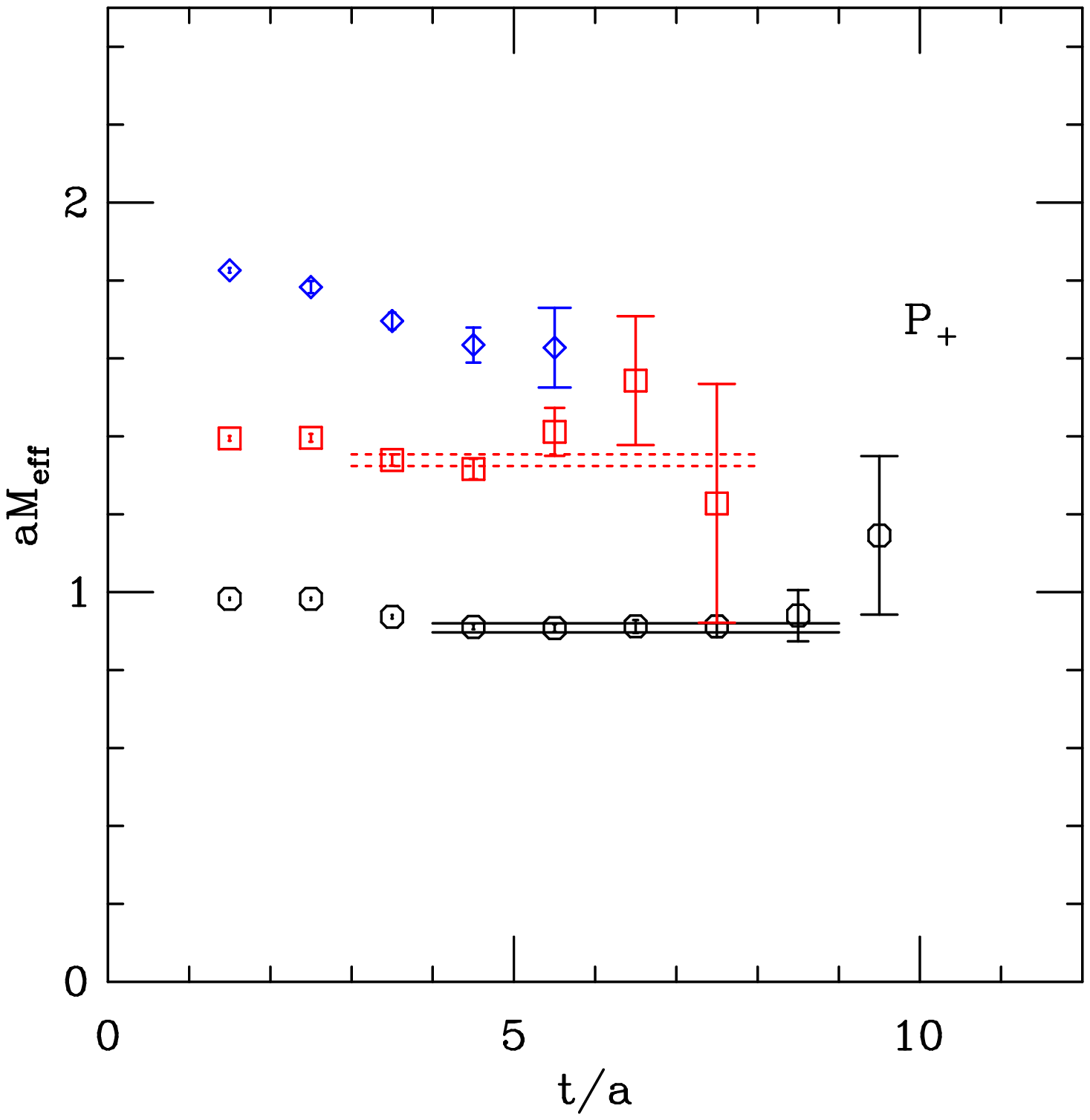}
\end{center}
\caption{
Effective masses for the static-light mesons on the quenched configurations. 
$am_q=0.08$, $a^{-1} \approx 1330$ MeV, $L \approx 1.8$ fm. 
The horizontal lines represent $M\pm\sigma_M$ fit values for the 
corresponding time ranges. 
Dashed lines indicate fits for which we adjust the minimum time for 
systematic error estimates.
}
\label{effmass}
\end{figure}

\begin{figure}
\begin{center}
\includegraphics*[width=6.5cm]{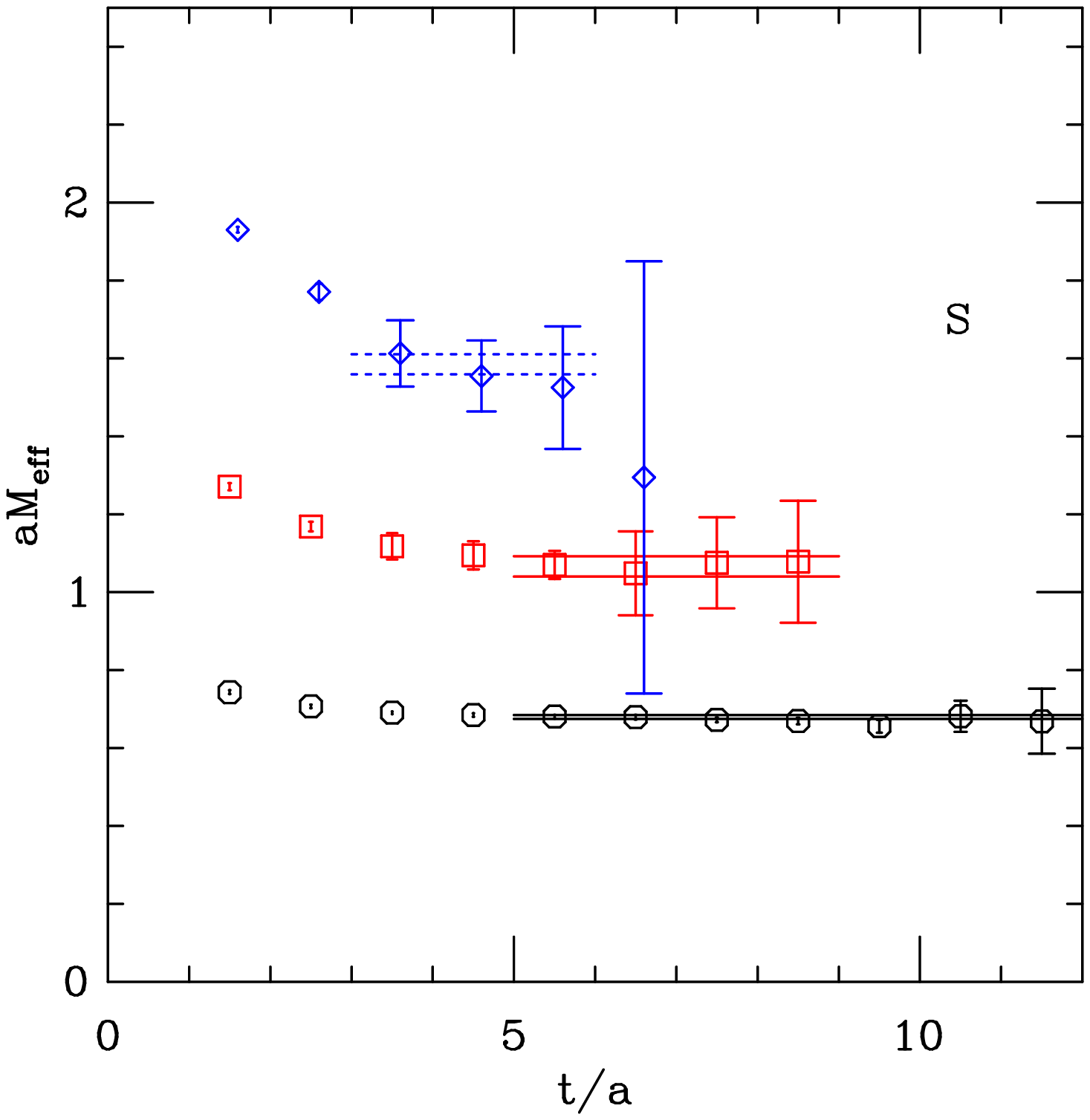}
\includegraphics*[width=6.5cm]{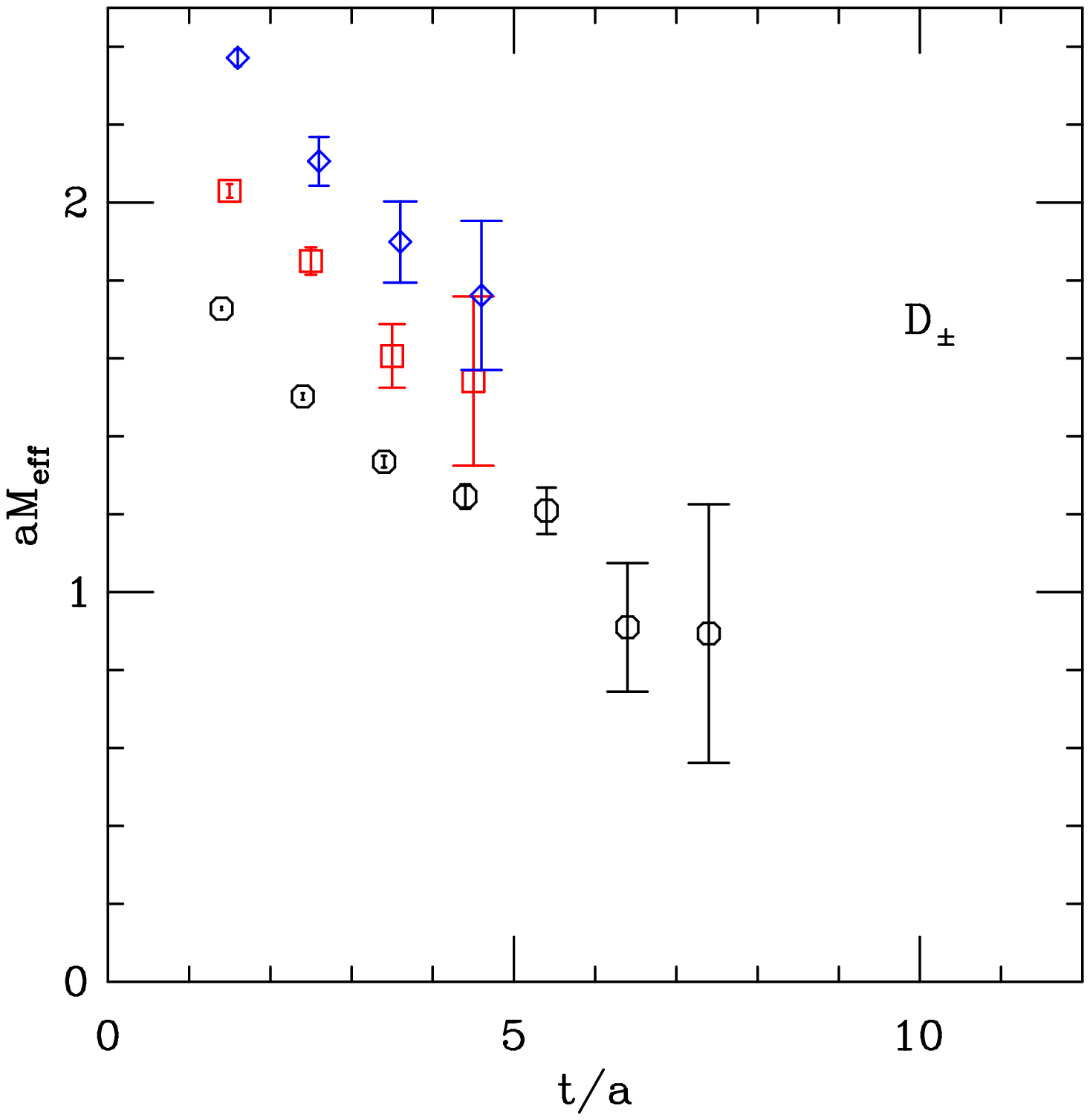}\\
\includegraphics*[width=6.5cm]{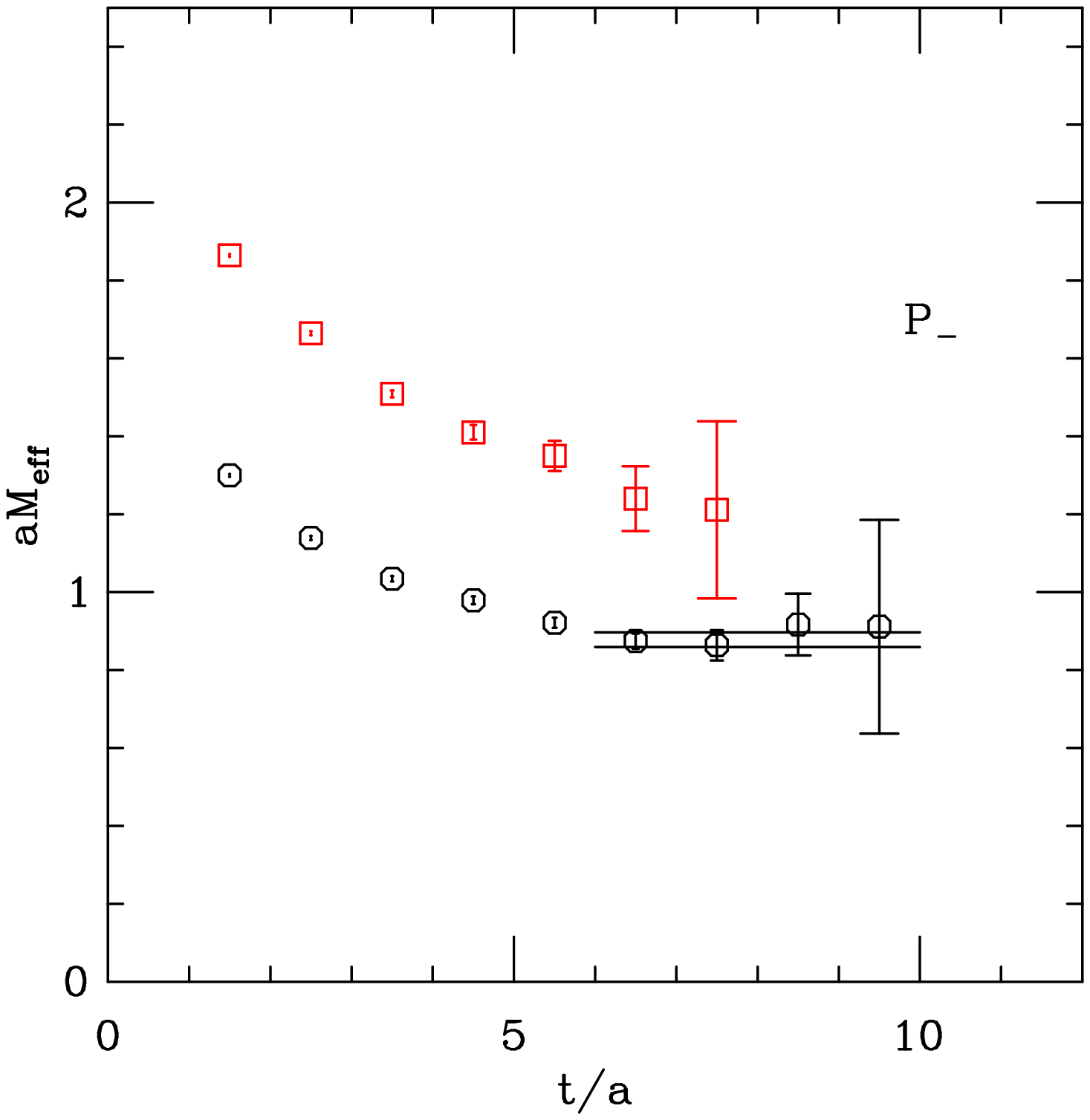}
\includegraphics*[width=6.5cm]{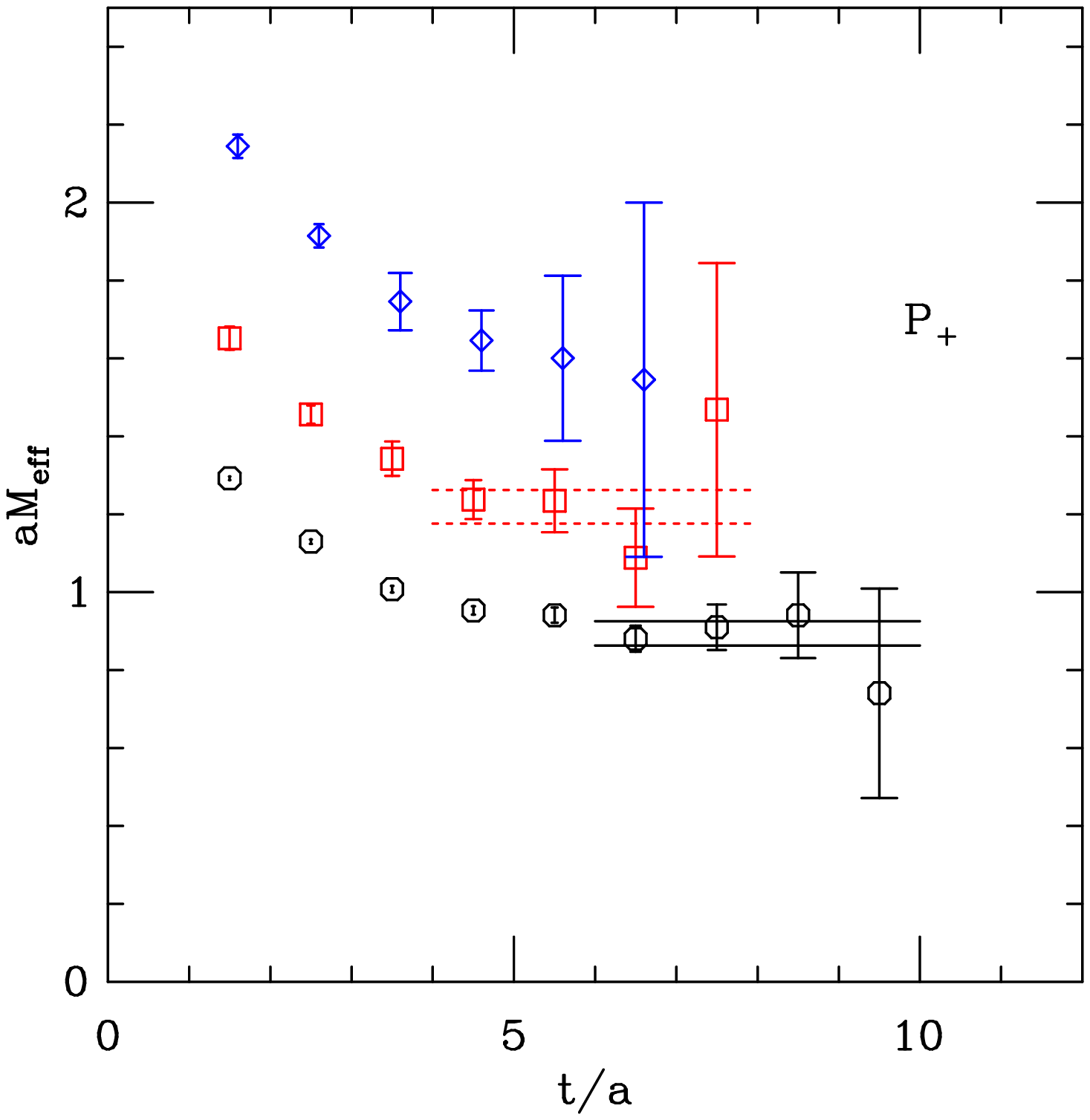}
\end{center}
\caption{
Effective masses for the static-light mesons on the dynamical 
($M_{\pi,\mbox{sea}} \approx 500$ MeV) configurations. 
$am_q=0.08$, $a^{-1} \approx 1710$ MeV, $L \approx 1.4$ fm.
The horizontal lines represent $M\pm\sigma_M$ fit values for the 
corresponding time ranges. 
Dashed lines indicate fits for which we adjust the minimum time for 
systematic error estimates.
}
\label{effmass_dyn}
\end{figure}

\begin{figure}
\begin{center}
\includegraphics*[width=8cm]{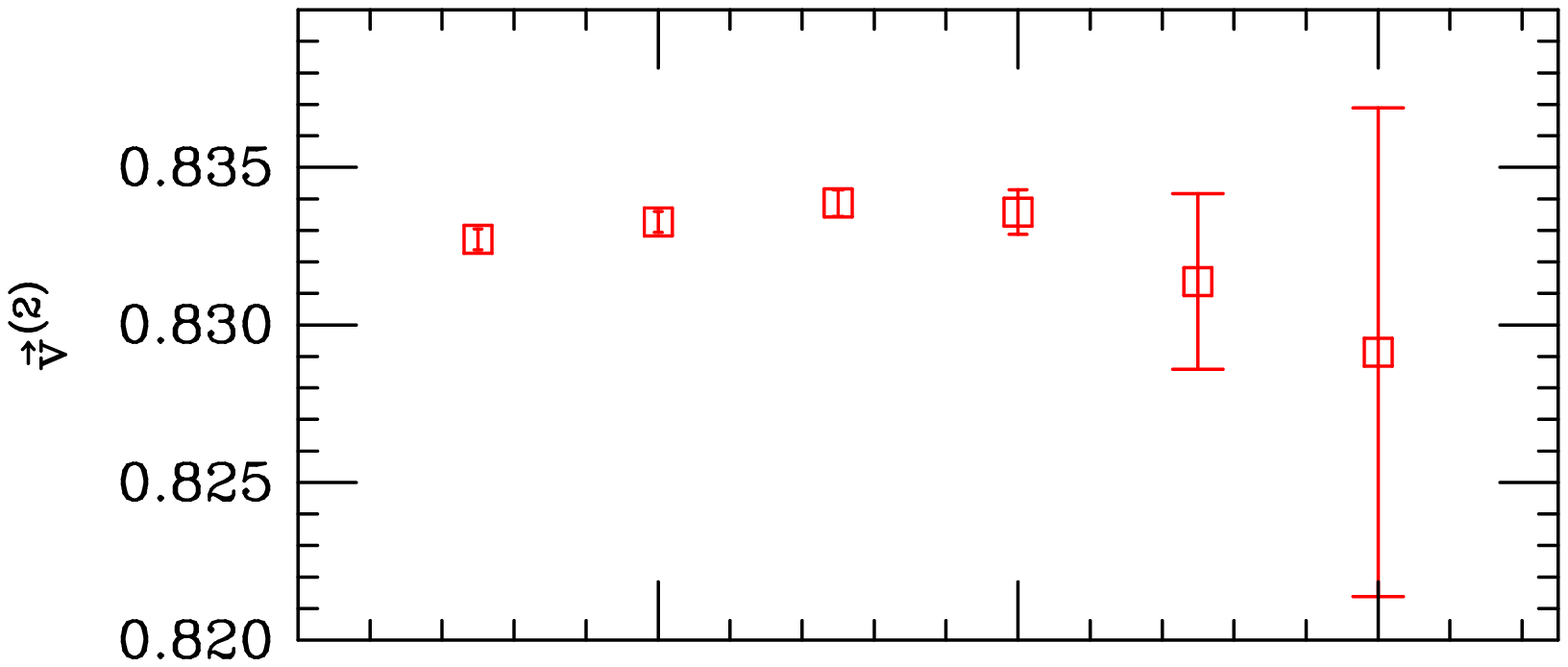}\\
\includegraphics*[width=8cm]{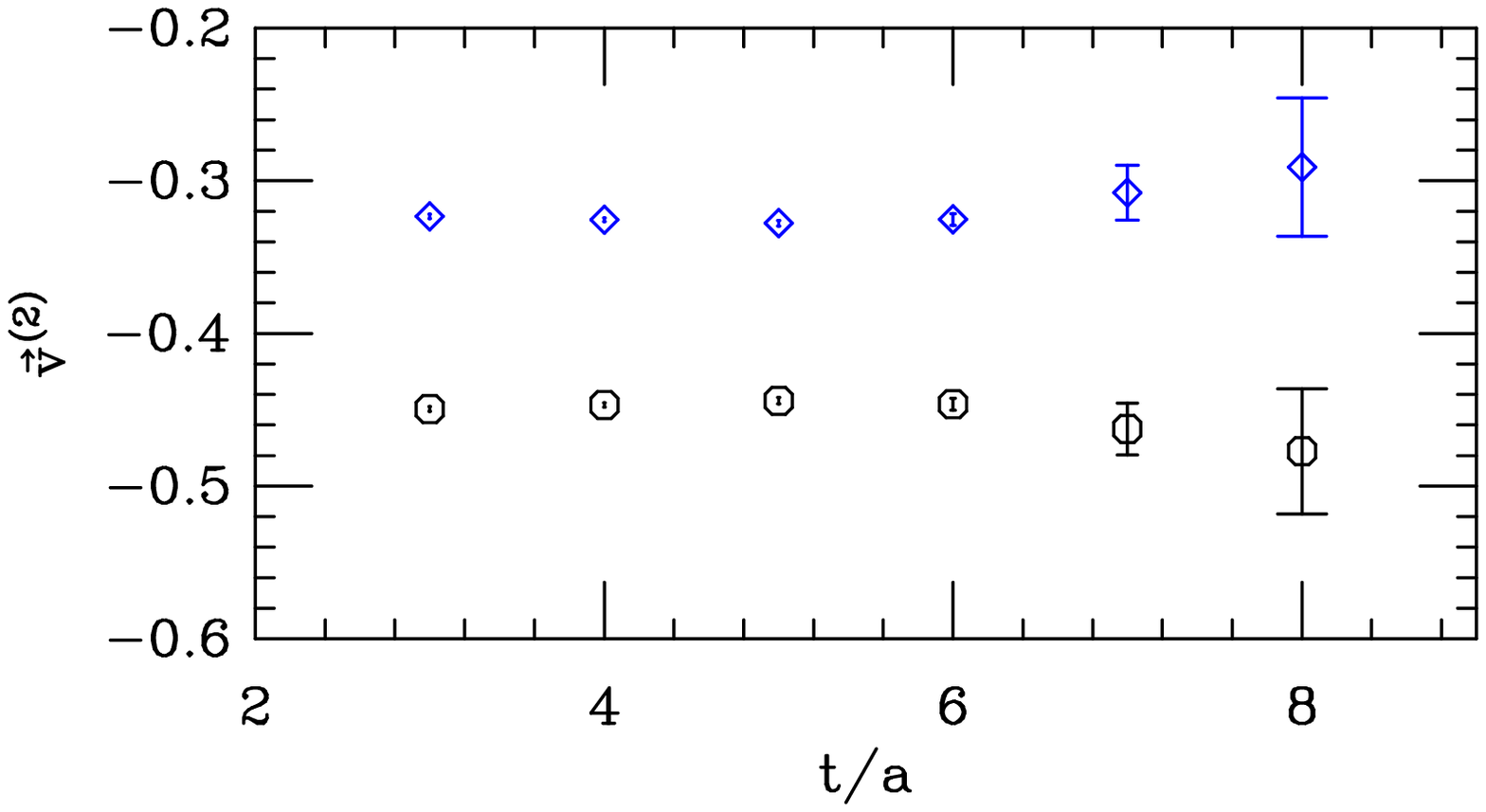}
\end{center}
\caption{
The three eigenvector components for the quenched $2P_+$ state over the time 
range of the fit.
}
\label{evec2Pp}
\end{figure}

After extracting the eigenvalues, we check for single mass states by 
creating effective masses:
\begin{equation}
  aM_{\mbox{eff}}^{(k)}\left(t + \frac{1}{2}\right) = 
  \ln \left( \frac{\lambda^{(k)}(t)}
    {\lambda^{(k)}(t+1)} \right) \, .
\end{equation}
A representative sample of these, along with single-elimination jackknife 
errors, are plotted against time in Figures \ref{effmass} and 
\ref{effmass_dyn}. 
In each case one finds values from the first two or three eigenvalues. 
The horizontal lines signify the $M\pm\sigma_M$ values which result from 
correlated fits over the corresponding range in time 
(dashed lines denote fits where we also later increased the minimum time of 
the fit in order to estimate systematic errors). 
We require that at least three effective mass points display a plateau 
(within errors) and that the eigenvectors remain constant over the same 
range before we perform said fits. 
Figure \ref{evec2Pp} shows the eigenvector components for the quenched 
first-excited $P$-wave ($2P_+$), a state displaying a rather jumpy 
effective-mass plateau. There is some slight variation in the central values 
over time, but when considering the fine scale of the plots and the errors, 
we are confident that we are dealing with a single state.

\begin{figure}
\begin{center}
\includegraphics*[width=6.5cm]{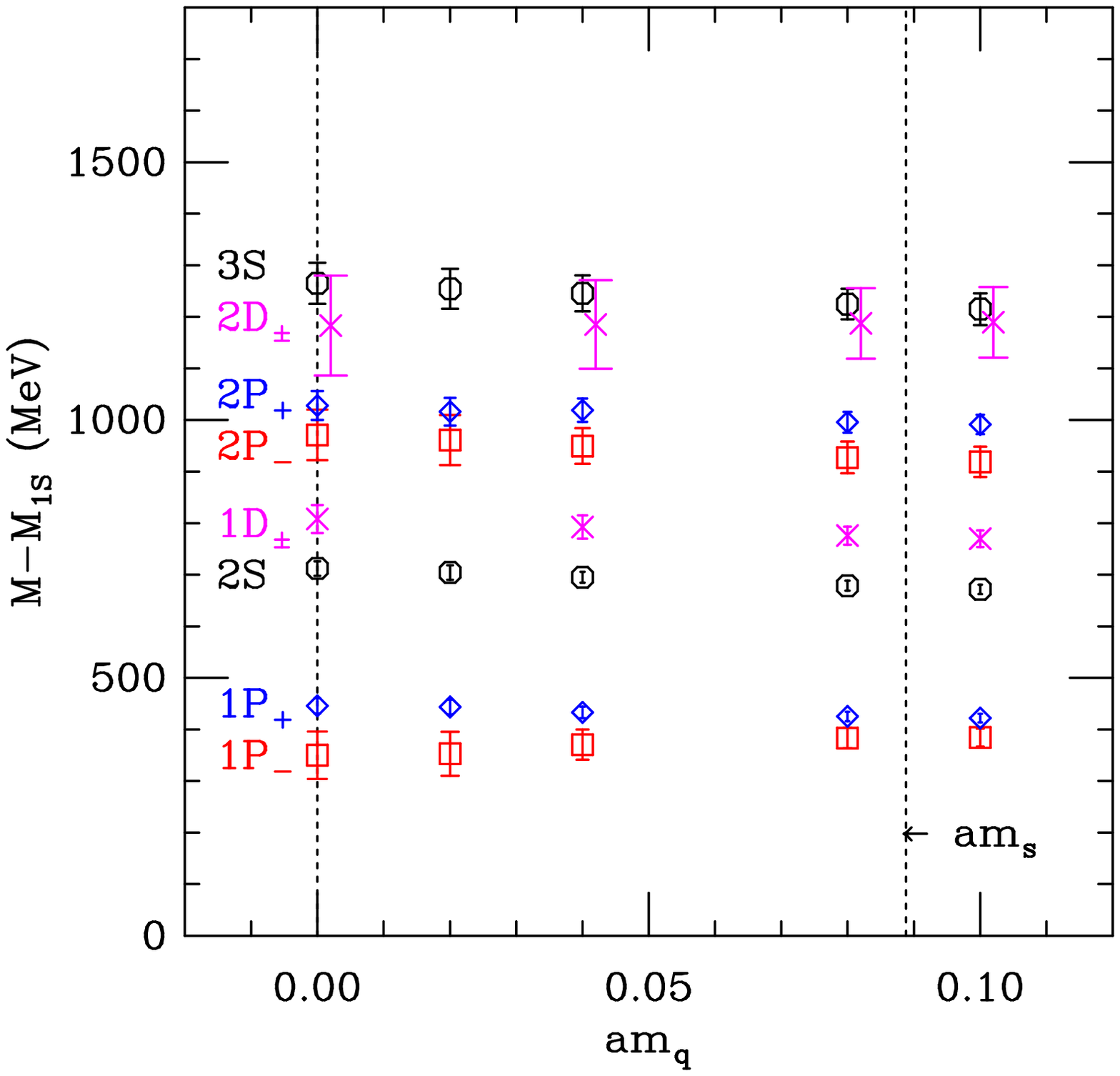}
\includegraphics*[width=6.5cm]{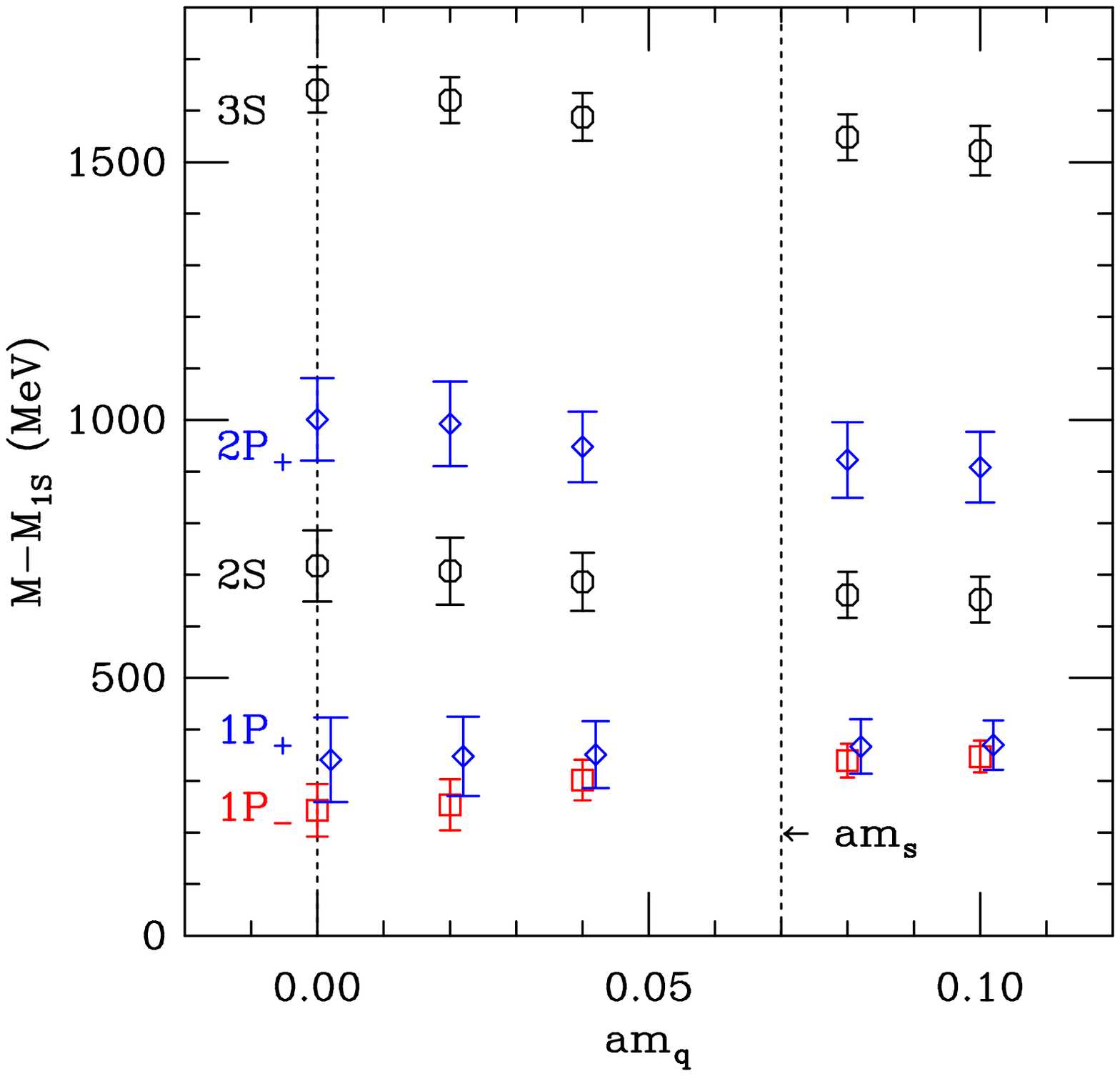}
\end{center}
\caption{
Physical mass splittings ($M-M_{1S}$) as a function of the quark mass for 
the quenched (left) and dynamical (right) lattices. 
The vertical lines denote the chiral limit ($m_q\to0$) and the strange quark 
mass ($m_s$). 
The left-most values are the (linear) chiral extrapolations. 
All errors are only statistical for the chosen fit ranges.
}
\label{chiral_extrap}
\end{figure}

Performing fits for all quark masses, we next take a look at the mass 
splittings ($M-M_{1S}$) as a function of the quark mass. 
These are plotted in Figure \ref{chiral_extrap}, along with the chirally 
extrapolated results ($m_q\to0$). We use simple linear fits to perform these 
extrapolations (as well as for the interpolations to $m_q=m_s$) since there 
appears to be little scaling in these plots. 
Both sets of results display a clear ordering of alternating parities with 
increasing mass. 
The quenched $2S$ and $1D_\pm$ states are close, with the latter being 
slightly higher in mass. 
This is expected since the $D_\pm$ represents an average of higher 
spins than the $S$ and the heavy-quark spin interactions which would ``mix'' 
the purely orbitally excited $D$-waves with the ground-state $S$-wave are 
absent in the static approximation. 
Overall, the results for the dynamical configurations are poorer. 
This is no big surprise since here we have not only fewer configurations, 
but also a smaller physical volume ($L\approx1.4$ fm). 
This fact also makes it difficult, at least at the present stage, to 
discern quenching and finite-volume effects. 
However, the one marked difference, the jump in the $3S$ mass, is quite 
likely due to the smaller volume.

In Tables \ref{B_split} and \ref{Bs_split} we present the results for the 
chirally extrapolated ($B$ mesons) and strange-quark-mass interpolated 
($B_s$ mesons) mass splittings. 
We include the statistical errors from the fits in the first set of 
parentheses. 
For fits where the effective-mass plateaus are not immediately clear 
(e.g., fits represented with dashed lines in Figures \ref{effmass} and 
\ref{effmass_dyn}), we move the minimum time of the fit range out by 
one to two time slices and observe the subsequent changes in 
$M\pm\sigma_M$, as compared to the previous values. 
The differences from the old values are reported as systematic errors; 
these appear in the second set of parentheses.

\begin{table}
\caption{
Mass splittings for our $B$ mesons ($m_q\to0$). 
The first number in parentheses is the statistical error. 
The second set (if present) are the additional systematic errors which result 
from adjustments to the minimum time of the fit.
\label{B_split}
}
\begin{tabular}{lcccc} \hline
state & $J^P$ & \multicolumn{3}{c}{$M-M_{1S}$ (MeV)} \\
 & & $N_f=0$, $L \approx 1.8$ fm & $N_f=2$, $L \approx 1.4$ fm & PDG \cite{PDG} \\ \hline
$2S$ & $0^-,1^-$ & 712(14) & 717(69) & - \\
$3S$ & $0^-,1^-$ & 1265(40)($^{+0}_{-130}$) & 1640(44)($^{+55}_{-200}$) & - \\ \hline
$1P_-$ & $0^+,1^+$ & 350(46) & 243(51) & 384(8) \\
$2P_-$ & $0^+,1^+$ & 971(49)($^{+50}_{-90}$) & - & - \\ \hline
$1P_+$ & $1^+,2^+$ & 446(15) & 341(82) & 384(8) \\
$2P_+$ & $1^+,2^+$ & 1028(28)($^{+160}_{-80}$) & 1001(80)($^{+130}_{-20}$) & - \\ \hline
$1D_\pm$ & $1^-,2^-,3^-$ & 808(27)($^{+0}_{-90}$) & - & - \\
$2D_\pm$ & $1^-,2^-,3^-$ & 1183(97)($^{+130}_{-150}$) & - & - \\ \hline
\end{tabular}
\end{table}

It is interesting to note that, in moving from the quenched lattices to 
the smaller, dynamical lattices, the $M_{2S}-M_{1S}$ mass splitting remains 
unchanged. 
Unless there is an odd counter-balancing of finite-volume and quenching 
effects between the lattices, we may conclude that our values for 
$M_{2S}-M_{1S}$ are reliable. 
The same is true for the $M_{2P}-M_{1S}$ splitting, which is a bit strange 
since $M_{1P}-M_{1S}$ already shows a difference, its value becoming 
especially low for the $1P_-$, when compared with the result from 
experiment (the quantum numbers, $J^P$, for the experimentally observed, 
excited $B$ and $B_s$ mesons have yet to be confirmed, so we place them in 
both the $1P_-$ and $1P_+$ rows). 
For our $B_s$ mesons, all $M_{1P}-M_{1S}$ splittings appear to be too small. 
Better statistics and larger volumes are needed if we are to clearly resolve 
these matters.

\begin{table}
\caption{
Mass splittings for our $B_s$ mesons ($m_q=m_s$). 
The first number in parentheses is the statistical error. 
The second set (if present) are the additional systematic errors which result 
from adjustments to the minimum time of the fit.
\label{Bs_split}
}
\begin{tabular}{lcccc} \hline
state & $J^P$ & \multicolumn{3}{c}{$M-M_{1S}$ (MeV)} \\
 & & $N_f=0$, $L \approx 1.8$ fm & $N_f=2$, $L \approx 1.4$ fm & PDG \cite{PDG} \\ \hline
$2S$ & $0^-,1^-$ & 675(10) & 665(45) & - \\
$3S$ & $0^-,1^-$ & 1220(30)($^{+20}_{-50}$) & 1560(45)($^{+35}_{-190}$) & - \\ \hline
$1P_-$ & $0^+,1^+$ & 384(20) & 330(34) & 448(16) \\
$2P_-$ & $0^+,1^+$ & 923(30)($^{+10}_{-60}$) & - & - \\ \hline
$1P_+$ & $1^+,2^+$ & 424(10) & 363(55) & 448(16) \\
$2P_+$ & $1^+,2^+$ & 993(20)($^{+130}_{-50}$) & 930(75)($^{+0}_{-80}$) & - \\ \hline
$1D_\pm$ & $1^-,2^-,3^-$ & 773(17)($^{+0}_{-80}$) & - & - \\
$2D_\pm$ & $1^-,2^-,3^-$ & 1188(68)($^{+170}_{-80}$) & - & - \\ \hline
\end{tabular}
\end{table}

One thing is clear though: due to the improvement of the light-quark 
propagator estimation, and our subsequent ability to use half the points of 
the lattice as source locations, we have greatly improved our chances of 
isolating excited heavy-light states. 
In an earlier study \cite{Burch:2004aa} of heavy-light mesons using wall 
sources on the same quenched configurations, we were barely able to see the 
$2S$ state, let alone the excited states in any other operator channel. 
Also, there we used NRQCD for the heavy quark; this should only boost the 
signals since the heavy quark can then ``explore'' more of the lattice 
through its kinetic term. 
It is obvious, however, that we have much better signals now since we are 
able to see excited states in every channel ($2S$, $3S$, $2P_-$, $2P_+$, and 
$2D_\pm$) on the quenched lattice.

\subsection{Disconnected correlators}

As a preliminary testing ground for our closed propagators, we take a look 
at some first results for the disconnected contributions to pseudoscalar 
($J^P=0^-$) meson correlators: 
\begin{equation}
  \label{disc_corr}
  D(t) = \sum_{t_0,\vec x,\vec y} 
  \mbox{Tr}(\gamma_5 P_{\vec x,t_0;\vec x,t_0})
  \mbox{Tr}(\gamma_5 P_{\vec y,t_0+t;\vec y,t_0+t}) \, .
\end{equation}

Again, we use 12 random spin-color sources (initially placed everywhere on 
the lattice), spin-dilute them, and perform inversions ($P\chi^n$) at a 
quark mass of $am_q=0.02$ on the quenched configurations. 
We then condition these ``naive'' estimates via Eq.\ (\ref{disc_estimator2}) 
using the central point and its nearest neighbors (NN) as region 1 
(the calculation of all the $M_{11}^{-1}$'s for this sized region on a single 
configuration takes less than a day on a PC).

\begin{figure}
\begin{center}
\includegraphics*[height=6cm]{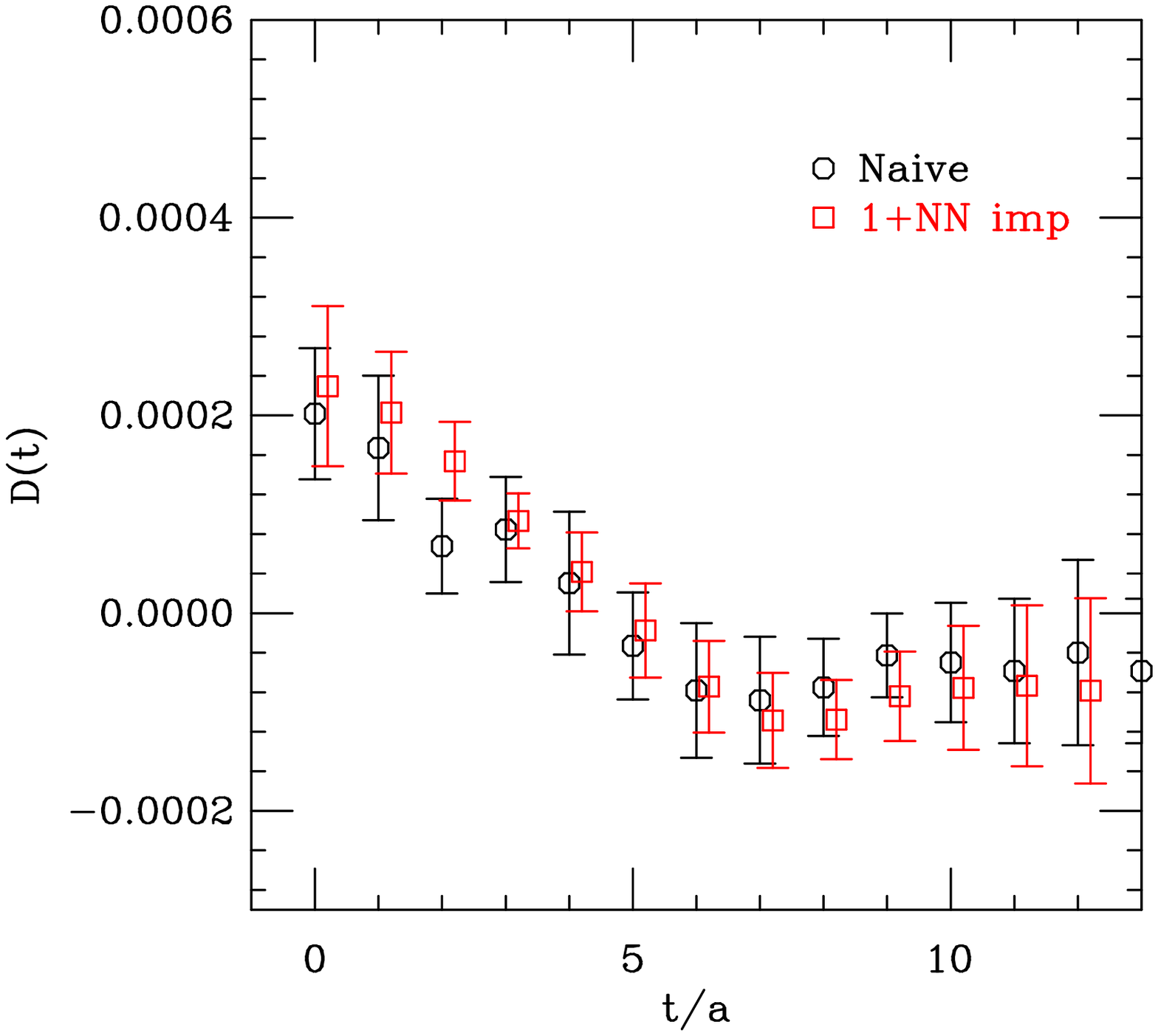}
\includegraphics*[height=6cm]{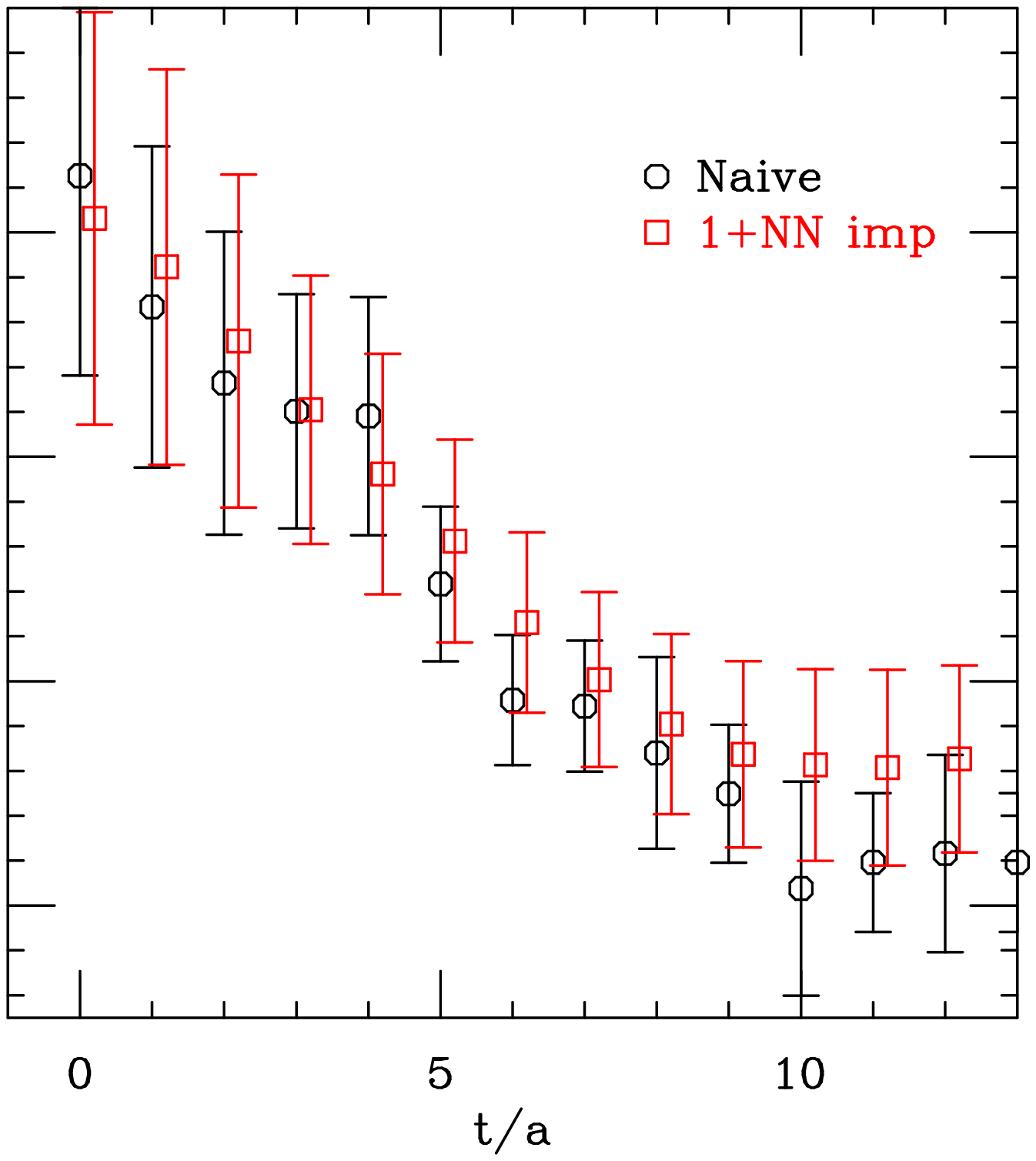}
\end{center}
\caption{
Naive and improved disconnected pseudoscalar correlators (with $am_q=0.02$) 
on two different quenched configurations. 
Errors result from a single-elimination jackknife procedure over the 12 
random spin-color sources.
}
\label{dg5_corr}
\end{figure}

In Figure \ref{dg5_corr} we compare results obtained via the naive and 
improved estimators on two different quenched configurations. 
The errors are estimated via the single-elimination jackknife subsets of 
the 12 random sources. 
Looking at the result for the configuration on the left, we can see 
significant reduction of the errors over many time separations. 
This is not the case for all configurations, however, as one can see on the 
right, where the errors are comparable, if not slightly larger for the 
improved version. 
For both cases shown here (and in fact for all configurations) the central 
values for the improved method follow a smoother curve. 
This should be no great surprise since the improved estimator uses sources 
on neighboring time slices (see Figure \ref{disc_sources}), whereas the 
naive one does not. 
So the improved version should smoothen out some of the remaining 
fluctuations over different $t$ values.

An important thing to note here is that, despite any of the improvement which 
we may gain from the smaller errors on some configurations and the 
smoothening of the curves, the error for the ensemble average of $D(t)$ 
will be dominated by the limited number of gauge configurations (i.e., this 
is a ``gauge-limited'' quantity). 
One can see from the figure that the fluctuation which comes from switching 
configurations is as big as, or bigger than, the jackknife errors from the 
sources. 
For some other configurations, the jump in the corresponding $D(t)$ values is 
much larger. 
These are perhaps configurations with large values of topological charge, 
$Q$; after all, the integrated disconnected pseudoscalar correlator is 
related to the square of this quantity \cite{Edwards:1998wx}: 
\begin{equation}
  \sum_t D(t) \propto \frac{Q^2}{m_q^2} \, ,
\end{equation}
where the relation is only approximate here since we use only chirally 
improved quarks.

Although we see no great overall improvement for these light, disconnected 
pseudoscalars, this is a tough testing ground. 
It remains to be seen how well these improved closed propagators will 
perform for other quantities where only averages of single loops over single 
time slices are needed: e.g., $s\bar s$ loops within hadron correlators to 
measure strangeness content, an application which we plan to look into in 
the near future. 
We also point the interested reader to other improvement schemes for closed 
contributions: see, e.g., \cite{McMi00,DeHe02}.

\section{Summary}

We have presented a method for improving the estimation of quark 
propagators in lattice QCD. 
The improvement we obtain for open propagators between equal subvolumes of 
the lattice is significant. 
The boost in statistics from being able to use up to half of the lattice 
sites as starting locations for our static-light meson correlators 
allows us to extract a number of radially and orbitally excited states 
($2S$, $3S$, $1P$, $2P$, $1D$, $2D$).

As we have pointed out, our method is similar to the maximal variance 
reduction approach \cite{Michael:1998sg}. 
However, since we can work with the Dirac matrix $M$, rather than 
$M^\dagger M$, our method may be better suited for highly improved actions.

Although the first results for our improved closed propagators are not very 
promising, it appears that we have chosen a very difficult application 
(light, disconnected, pseudoscalar correlators) for which to test them. 
We reserve final judgment on their usefulness for the future, when we 
measure their performance in other physical quantities.

\begin{ack}
We would like to thank Christof Gattringer for many helpful discussions. 
We are also indebted to our colleagues in Graz -- Christian B.\ Lang, 
Wolfgang Ortner, and Pushan Majumdar -- for sharing their dynamical CI 
configurations with us. 
We also wish to thank Andreas Sch\"afer for fostering a large research group, 
thereby making collaborations such as this possible. 
Simulations were performed on the Hitachi SR8000 at the LRZ in Munich. 
This work is supported by GSI.
\end{ack}

\section{Appendix A}

In Section \ref{SectMethod}, we derived the exact relations 
(\ref{postdiluting}) and (\ref{postdiluting2}) for the 
blocks of the quark propagator. In those derivations we relied upon the 
following property of the random vectors $\chi^{n}_i, \; n=1,\ldots,N$:
\begin{eqnarray}
\lim_{N\to\infty} \left( \frac{1}{N} \chi^{n}_i  \chi^{n\dagger}_j \right) & = & \delta_{ij}.
\end{eqnarray}
But there also exist other possibilities to derive the same results.

As an alternative, we show here how one can arrive at 
Eq.\ (\ref{postdiluting}) using the path integral over the fermionic 
degrees of freedom.

The Dirac propagator, $P=M^{-1}$, can be represented by a path integral in 
the following way 
\begin{eqnarray}
\label{fermpath}
P_{ij} & = & \frac{1}{Z} \int [D\psi] [D\bar{\psi}] \psi_i \bar{\psi}_j e^{-S_{f}} \, ,
\end{eqnarray}
where 
\begin{eqnarray}
\label{fermpartition}
Z & = & \int [D\psi] [D\bar{\psi}] e^{-S_{f}} \, ,
\end{eqnarray}
is the partition function, $S_{f}$ is the fermion action 
\begin{eqnarray}
S_{f} & = & \bar{\psi} M \psi \, ,
\end{eqnarray}
and $[D\psi] = \prod_k d\psi_k$ is the integration measure.

The first step in our alternative approach is to split the fermion action 
into a sum of four separate contributions:
\begin{eqnarray}
\label{splitaction}
S_{f} & = & \bar{\psi}_1 M_{11} \psi_{1} + \bar{\psi}_2 M_{22} \psi_{2}
+ \bar{\psi}_1 M_{12} \psi_{2} + \bar{\psi}_2 M_{21} \psi_{1}.
\end{eqnarray}
The leading two contributions are only connecting the fermion fields within
the regions $1$ ($R_1$) and $2$ ($R_2$), while the other two contributions 
provide the connections between the regions. 
In a similar way, one can also divide the integration measures in 
Eqs.\ (\ref{fermpath}) and (\ref{fermpartition}).

When one now considers the special case of Eq.\ (\ref{fermpath}) 
when $i \in R_1$ and $j \in R_2$ and uses expression (\ref{splitaction}) 
the propagator becomes 
\begin{eqnarray}
P_{12} & = & \frac{1}{Z} \int [D\psi] [D\bar{\psi}] \psi_1 \bar{\psi}_2 \,e^{-S_{f}} \nonumber \\
& = & - \frac{1}{Z} \int [D\psi_2] [D\bar{\psi}_2] \bar{\psi}_2 \,e^{-\bar{\psi}_2 M_{22} \psi_{2}} \times \nonumber \\
 & & \;\; \int [D\psi_1] [D\bar{\psi}_1] \psi_1 \, e^{-\bar{\psi}_1 M_{11} \psi_{1} - \bar{\psi}_1 M_{12} \psi_{2} - \bar{\psi}_2 M_{21} \psi_{1}} \nonumber \\
& = & - \frac{det(M_{11})}{Z} \int [D\psi_2] [D\bar{\psi}_2] \bar{\psi}_2 ( - M^{-1}_{11} M_{12} \psi_{2}) \,e^{-\bar{\psi}_2 \tilde{M}_{22} \psi_{2}} \nonumber \\
& = & - \frac{det(M_{11}) det(\tilde{M}_{22})}{Z} M^{-1}_{11} M_{12} \tilde{M}^{-1}_{22} \, ,
\end{eqnarray}
where $M_{11}$ and $\tilde{M}_{22} := M_{22} - M_{21} M^{-1}_{11} M_{12}$ 
are the diagonal blocks of a decomposed version of $M$ with 0's as 
off-diagonal blocks. 
Therefore, $Z=det(M)=det(M_{11}) det(\tilde{M}_{22})$ and the ratio 
becomes 1. 
For the next to last step, we completed the square in the exponent and 
performed a shift of the integration measure.

Since $\tilde{M}_{22}$ is the Schur complement of $M_{11}$, one can show that 
in fact $P_{22} = \tilde{M}^{-1}_{22}$. This leads us to our final result 
\begin{eqnarray}
P_{12} & = & - M^{-1}_{11} M_{12} P_{22} \, .
\end{eqnarray}


\end{document}